\documentclass{elsart}
\usepackage{amsmath,amssymb}
\newcommand{\1}{{\sf 1 \!\! 1}}

\newcommand{\diag}{\mbox{diag}}
\newcommand{\Tr}{\mbox{Tr}}
\newcommand{\sgn}{\mbox{sgn}}
\newcommand{\Sdet}{\mbox{Sdet}}
\newcommand{\oh}[1]{\overline{\widehat{#1}}}
\newcommand{\wh}[1]{\widehat{#1}}
\newcommand{\mvec}[2]{\left(\begin{array}{c}#1\\ #2\end{array}\right)}

\newcommand{\mat}[4]{\left(%
\begin{array}{cc}#1&#2 \\#3&#4\end{array}\right)}

\newcommand{\bra}[1]{\left\langle#1\right|}
\newcommand{\ket}[1]{\left|#1\right\rangle}

\newcommand{\braopket}[3]{\left\langle#1\left|#2\right|#3\right\rangle}

\DeclareMathOperator{\re}{Re}
\DeclareMathOperator{\im}{Im}

\DeclareMathOperator{\vol}{vol}

\begin{document}

\begin{frontmatter}
 
\title{Color-Flavor Transformation for the Special Unitary Group}
\author[Yale]{B. Schlittgen} and
\author[Yale,RBRC]{T. Wettig}
\address[Yale]{Department of Physics, Yale University, 
New Haven, CT 06520-8120} 
\address[RBRC]{RIKEN-BNL Research Center, Upton, NY, 11973-5000}

\date{5 February 2002}

\begin{abstract} 
  We extend Zirnbauer's color-flavor transformation in the fermionic
  sector to the case of the special unitary group. The transformation
  allows a certain integral over SU($N_c$) color matrices to be
  transformed into an integral over flavor matrices which parameterize
  the coset space U$(2N_f)/$U$(N_f)\times$U$(N_f)$. Integrals of the
  type considered appear, for example, in the partition function of
  lattice gauge theory.
\end{abstract}

\end{frontmatter}

\section{Introduction}

In the context of models describing disordered systems in condensed
matter physics, Zirnbauer recently \cite{Zirnbauer:1996} developed a
generalized supersymmetric Hubbard-Stratonovich transformation which
transforms a certain integral over U($N_c$) matrices into an integral
over matrices in the coset space
U$(n_{+}+n_{-}|n_{+}+n_{-})/$U$(n_{+}|n_{+})\times$U$(n_{-}|n_{-})$.
This transformation reads
\begin{multline}
  \label{eq:CFT}
\int_{\text{U}(N_c)}dU \exp\left( \bar{\psi}_{+a}^{i}U^{ij}\psi_{+a}^{j}+
\bar{\psi}_{-b}^{j}\bar{U}^{ij}\psi_{-b}^{i}\right) \\
=\int D\mu_{N_c}(Z,\tilde{Z})\exp\left(\bar{\psi}_{+a}^{i}Z_{ab}\psi_{-b}^i
+\bar{\psi}_{-b}^{j}\tilde{Z}_{ba}\psi_{+a}^{j} \right)\:.
\end{multline}
Here, the $\psi$-fields are $\mathbb{Z}_2$-graded tensors containing
bosonic and fermionic variables.  Integrals of this type (with only
fermionic degrees of freedom) appear in lattice gauge theory, where
$U$ represents a link variable.  It is therefore natural to think of
the indices $i,j=1,\ldots,N_c$ as color, and of the indices
$a=(\alpha,\sigma)$ and $b=(\beta,\sigma)$ as flavor, where
$\alpha=1,\ldots,n_{+}$, $\beta = 1,\ldots,n_{-}$, and bosonic and
fermionic components are denoted by $\sigma = B,F$.  The bar denotes
complex conjugation.

The transformation \eqref{eq:CFT} trades the integral over the gauge
fields $U$, which couple the color degrees of freedom of $\psi$, for
an integral over color-singlet fields $Z$ and $\tilde Z$, which
couple the flavor degrees of freedom of $\psi$.  The matrices $Z$ and
$\tilde{Z}$ on the right-hand side are supermatrices whose boson-boson
(fermion-fermion) blocks satisfy the relation
$\tilde{Z}_{BB}=Z^{\dag}_{BB}$ ($\tilde{Z}_{FF}=-Z^{\dag}_{FF}$).  The
integration measure is given by
\begin{equation}
  d\mu_{N_c}(Z,\tilde{Z})=D(Z,\tilde{Z})\Sdet(1-\tilde{Z}Z)^{N_c}\:,
\end{equation}
where $D(Z,\tilde{Z})$ is the flat Berezin measure and $\Sdet$ denotes
the super determinant. Furthermore, integration is constrained to the
region in which all eigenvalues of $\tilde{Z}_{BB}Z_{BB}$ are less
than unity.

The color-flavor transformation has been applied to derive a field
theory of the random flux model \cite{Altl99} and, in the context of
lattice gauge theories, to derive chiral Lagrangians in the
strong-coupling and large-$N_c$ limits 
\cite{Budczies:2000a,Nagao:2000ke,Budczies:2000qs}.  The extension of
the transformation to the cases of gauge group Sp($2N_c$) and O($N_c$)
is relatively straightforward.  It was worked out in Refs.\ 
\cite{Zirnbauer:1996,Zirnbauer:1997qm} and applied to lattice gauge
theories in Ref.~\cite{Nagao:2000ke}.  However, we would also like to be 
able to apply
the transformation to the lattice version of quantum chromodynamics
(QCD) where the gauge group is SU($N_c$) (with $N_c=3$) rather than
U($N_c$).  This case turns out to be more difficult.  Since our main
interest is in lattice gauge theory with fermionic matter fields, we
restrict ourselves for now to the fermionic sector, 
in which the tensor $\psi$
contains only anticommuting components.  A first step towards extending
the transformation to the special unitary group
was taken in Ref.~\cite{Budczies:2000qs} where it was
realized that the right-hand side of Eq.~\eqref{eq:CFT} becomes a
sum over disconnected sectors, each characterized by a different U(1)
charge $Q$, where $Q=-N_f,\ldots, N_f$.  However, in that paper,
only the terms
corresponding to $Q=0$ and 1 were given, without proof.  Moreover,
these terms involved unspecified constants.  

In the present paper, we derive the complete color-flavor
transformation for SU($N_c$). Applications to lattice gauge theory are
deferred to future work.  It should be noted that the integral over
the gauge group in Eq.~\eqref{eq:CFT} does not contain the gauge
action but only the terms coupling quarks to gluons.  For applications
of the transformation beyond the strong-coupling limit it will be
essential to find a way of including the gauge action in the integral.

This paper is organized as follows.  In Sec.~\ref{sec:result},
we state our result for the convenience of the reader who is less
interested in the technical details.  Sec.~\ref{sec:prelim} collects a
number of mathematical preliminaries that are needed for the proof of
our result in Sec.~\ref{sec:proof}.  Conclusions are drawn and an
outlook on possible applications of the color-flavor transformation in
lattice gauge theory and generalizations to the case involving both
bosonic and fermionic matter fields is given in Sec.~\ref{sec:concl}.
Finally, the appendix contains some simple examples to illustrate
the transformation.

\section{Statement of the result}
\label{sec:result}

Our result for the color-flavor transformation involving the special
unitary group is
\begin{align}
\label{eq:col_flav}
\int_{\text{SU}(N_c)} & dU \exp\left(\bar{\psi}_{a}^{i}U^{ij}\psi_{a}^{j}+
\bar{\varphi}_{a}^{i}U^{\dagger ij}\varphi_{a}^{j} \right)\notag\\ 
&=C_0\int_{\mathbb{C}^{N_f\times N_f}} \frac{D(Z,Z^{\dag})}
{\det(1+ZZ^{\dag})^{N_c}}
\exp\left(\bar{\psi}_a^i Z_{ab}\varphi_b^i
-\bar{\varphi}_a^iZ_{ab}^{\dag}\psi_b^i \right)
\sum_{Q=0}^{N_f} \chi_Q \:,
\intertext{where}
\chi_0=&1\:, \qquad\chi_{Q>0}=\mathcal{C}_Q
\left( \det(\mathcal{M})^Q+\det(\mathcal{N})^Q\right)\:.
\label{eq:chi_q}
\end{align}
Here, $\psi$ and $\varphi$ are tensors containing only fermionic
variables, with color indices $i,j=1,\ldots,N_c$ and flavor indices
$a,b=1,\ldots,N_f$.  The entries of the $N_c\times N_c$ matrices
$\mathcal{M}$ and $\mathcal{N}$ are given by
$\mathcal{M}^{ij}=\bar{\psi}_a^i(1+ZZ^{\dag})_{ab}\psi_b^j$ and
$\mathcal{N}^{ij}=\bar{\varphi}_a^i(1+Z^{\dag}Z)_{ab}\varphi_b^j$.
The overall normalization constant $C_0$ is consistent with
$\vol(\text{SU}(N_c))=1$, see Eqs.~\eqref{eq:norm} and
\eqref{eq:norm2}.  The constants $\mathcal{C}_Q$ are given by
\begin{equation}
  \label{eq:CQ}
  \mathcal{C}_Q = \frac{1}{(Q!)^{N_c} (N_c!)^{Q}}\prod_{n=0}^{Q-1}
  \frac{(N_c+n)!(N_f+n)!}{n!(N_c+N_f+n)!}\:.
\end{equation}
The matrix $Z$ is a complex matrix of dimension
$N_f$, and the measure of integration in Eq.~\eqref{eq:col_flav} 
reads
\begin{equation}
  \label{eq:measure}
  D(Z,Z^{\dag})=C\:\frac{dZ
  dZ^{\dag}}{\det(1+ZZ^{\dag})^{2N_f}}\quad\text{with}\quad
  dZdZ^\dagger=\prod_{a,b=1}^{N_f}d\re Z_{ab}\;d\im Z_{ab}
\end{equation}
and a constant $C$ chosen so that $\int D(Z,Z^\dagger)=1$, see
Eq.~\eqref{eq:normNf}.

The next two sections are dedicated to the proof of
Eq.~\eqref{eq:col_flav}.  To a large
extent, we shall proceed along the lines of Zirnbauer's original
proof in Ref.~\cite{Zirnbauer:1996}, but several extensions are needed
to adapt it to the case of SU($N_c$).  We shall give a
complete account of all relevant details to make the discussion
self-contained.

Let us note that the above result can easily be generalized to the
case where the tensors $\psi$ and $\varphi$ contain unequal numbers
of flavors, i.e., the flavor index on $\psi$ ($\varphi$) runs from 1
to $N_f^1$ ($N_f^2$).  In this case, the matrix $Z$ has dimension
$N_f^1\times N_f^2$, and the sum over $Q$ in Eq.~\eqref{eq:col_flav}
is replaced by
\begin{equation}
  1+\sum_{Q=1}^{N_f^1}\mathcal{C}_Q^1\det(\mathcal{M})^Q+
  \sum_{Q=1}^{N_f^2}\mathcal{C}_Q^2\det(\mathcal{N})^Q\:,
\end{equation}
with $\mathcal{C}_Q^1$ ($\mathcal{C}_Q^2$) given by the
$\mathcal{C}_Q$ from Eq.~\eqref{eq:CQ} with $N_f$ replaced by $N_f^2$
($N_f^1$).

\section{Mathematical preliminaries}
\label{sec:prelim}
In this section, we introduce the mathematical framework needed to
prove Eq.~\eqref{eq:col_flav} and quote some of the relevant
intermediate results. This will allow us to give the proof in
Sec.~\ref{sec:proof} in a relatively straightforward manner.

\subsection{Generalized coherent states and projection onto the {\rm
    SU}($N_c$) neutral sector}
\label{coherent}
We introduce two sets of fermionic creation and annihilation
operators, $\bar{c}_a^i$, $c_a^i$ and
$\bar{d}_a^i$, $d_a^i$, where in each
case, $a$ runs from 1 to $N_f$, and $i$ runs from 1 to $N_c$.  
These operators may be interpreted as quantized versions of 
the classical Grassmann
variables $\varphi_a^i$ and $\psi_a^i$, respectively.
For simplicity, we also introduce an index $A=1,\ldots,2N_f$ such that
\begin{equation}
  c_A^i=\left\{
    \begin{array}{ll}
      c_A^i & \text{for\ } A\le N_f\:, \\
      d_{A-N_f}^i & \text{for\ } A>N_f
    \end{array}\right.
\end{equation}
and similarly for $\bar c_A^i$.  In general, we shall use lower case
letters $a,b$ etc.\ to denote indices ranging from 1 to $N_f$, and
upper case letters $A,B$ etc.\ to denote indices ranging from 1 to
$2N_f$. The creation and annihilation operators satisfy canonical
anticommutation relations,
\begin{equation}
\{c_a^i,\bar{c}_b^j\}=\{d_a^i,\bar{d}_b^j\}=\delta^{ij}\delta_{ab}\:.
\end{equation}
We also define a Fock vacuum $\ket{0}$ which is annihilated by both $c_a^i$
and $d_a^i$, i.e.\ $c_a^i\ket{0}=d_a^i\ket{0}=0$. The Fock space is
generated by acting on $\ket{0}$ with all possible combinations of
creation operators $\bar{c}_a^i$ and $\bar{d}_a^i$. In the dual space,
we correspondingly have a vacuum state $\bra{0}$ with the property
$\bra{0}\bar{c}_{a}^i=\bra{0}\bar{d}_a^i=0$.

The set of all bilinear products of one creation operator with one 
annihilation operator forms a set of generators of a gl$(2 N_f N_c)$
algebra,
\begin{equation}
  \begin{array}{l@{\hspace{10mm}}l}
    E_{ab}^{ij}\equiv \bar{c}_a^i c_{b}^j -
    \frac{1}{2}\delta_{ab}\delta^{ij}\:,& 
    E_{a,N_f+b}^{ij}\equiv \bar{c}_a^i d_b^j\:,\\
    E_{N_f+a,b}^{ij}\equiv \bar{d}_a^i c_b^j\:, &
    E_{N_f+a,N_f+b}^{ij}\equiv \bar{d}_a^i
    d_b^j-\frac{1}{2}\delta_{ab}\delta^{ij}\:, 
  \end{array}
\end{equation}
where $a,b=1,\ldots,N_f$ and $i,j=1,\ldots,N_c$. For convenience, we have
chosen these operators not to be traceless. They satisfy
the usual commutation relations,
\begin{equation}
  [E_{AB}^{ij},E_{CD}^{k\ell}]=\delta_{CB}\delta^{kj}E_{AD}^{i\ell}
  -\delta_{AD}\delta^{i\ell}E_{CB}^{kj}\:.
\end{equation}

This algebra has two subalgebras that are important for our purposes,
namely gl$(2N_f)$, which is generated by $\{\sum_{i=1}^{N_c}E_{AB}^{ii}\}$,
and sl($N_c$), which is generated by $\{\mathcal{E}^{ij}\equiv
\sum_{A=1}^{2N_f}E_{AA}^{ij};i\neq j\}$ and 
$\{\mathcal{H}^i\equiv\sum_{A=1}^{2N_f}(E_{AA}^{ii}-E_{AA}^{N_c N_c}); 
i=1,\ldots, N_c-1\}$. These two subalgebras commute.

An essential part of Zirnbauer's proof is the projection of states
in the Fock space onto the color-neutral sector.  This is the subset
of Fock space whose elements are invariant under (in our case)
SU($N_c$) rotations, i.e., for a color-neutral state
$\ket{\mathcal{N}}$, we have
\begin{equation}
  \label{eq:cns}
  \mathcal{E}^{ij}\ket{\mathcal{N}}=0\quad\text{and}\quad
  \mathcal{H}^i\ket{\mathcal{N}}=0
\end{equation}
for all generators in the sl($N_c$) subalgebra.  The basic idea of
the proof is to derive two different implementations of a projection
operator onto the color-neutral sector.  One of these is obtained by
integrating over the gauge group, corresponding to the left-hand side
of Eq.~\eqref{eq:col_flav} (see Sec.~\ref{fermi}).  The other one,
corresponding to the right-hand side of Eq.~\eqref{eq:col_flav}, is
obtained using some properties of generalized coherent states and
integrating over a certain coset space $G/H$ (see this and the
next subsection).  Identification of the two projection
operations then establishes Eq.~\eqref{eq:col_flav}.

Equation~\eqref{eq:cns} may equivalently be written as
\begin{equation}
  \label{eq:color-neutral}
  \sum_{a=1}^{N_f}\left( \bar{c}_a^i c_a^j+\bar{d}_a^i d_a^j \right)
  \ket{\mathcal{N}} =(N_f+Q)\delta^{ij} \ket{\mathcal{N}}\:.  
\end{equation}
Here, $Q$ can take on integer values between $-N_f$ and $N_f$.  These
bounds on $Q$ are due to the fact that the Fock space contains only
fermions, and the operator on the left-hand side of
Eq.~\eqref{eq:color-neutral} simply counts the number of fermions in
the state $\ket{\mathcal{N}}$.  (Note also that this is the place
where the differences to the original case of U($N_c$) start to show
up.  In the latter case, the color-neutral sector is invariant under
U($N_c$), and there is an additional diagonal generator
$\mathcal{H}^{N_c}$ which annihilates color-neutral states.  This
implies that $Q=0$ in Eq.~\eqref{eq:color-neutral} so that there is
no sum over $Q$ in Eq.~\eqref{eq:col_flav}.)

In order to derive the projector onto the color-neutral sector, we
consider the action of the group U($2N_f$) on the Fock space.  (The
restriction of Gl($2N_f$) to the submanifold U($2N_f$) leads to a
projection onto the SU($N_c$), rather than Sl($N_c$), invariant
sector, and it leads to a Riemannian integration domain on the
right-hand side of Eq.~\eqref{eq:col_flav}.)  In the following, we
show how to associate to each value of $Q$ a particular irreducible
representation of U($2N_f)$.

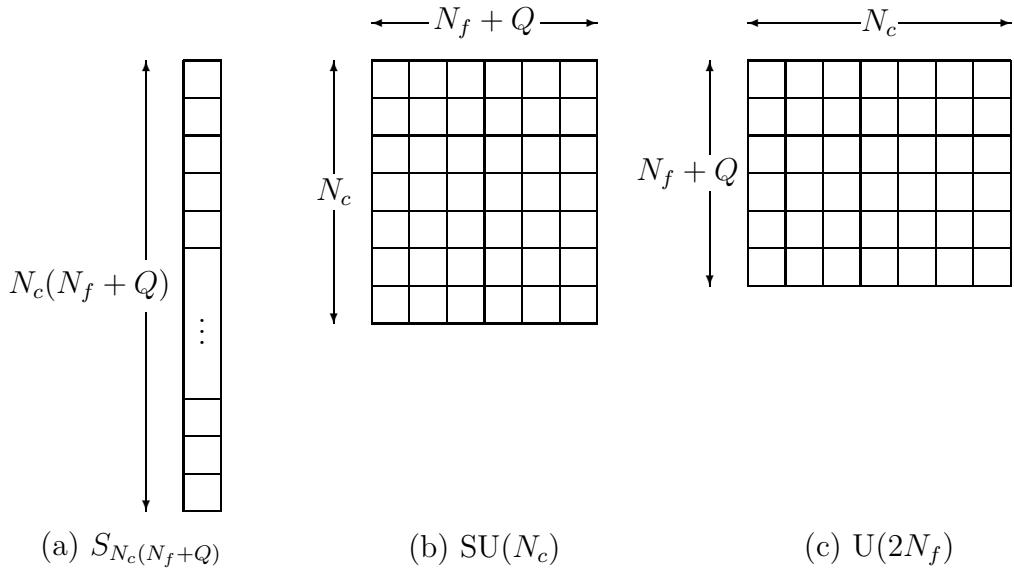
\begin{figure}[tb]
\begin{center}
\unitlength1cm
\begin{picture}(12.5,7.0)

\put(1.5,3.85){\vector(0,1){2.65}}
\put(1.5,3.15){\vector(0,-1){2.65}}

\put(0.75,3.5){\makebox(0,0){$N_c(N_f+Q)$}}

\put(2,0.5){\line(1,0){0.5}}
\put(2,1.0){\line(1,0){0.5}}
\put(2,1.5){\line(1,0){0.5}}
\put(2,2.0){\line(1,0){0.5}}
\put(2.25,3.0){\makebox(0,0){$\vdots$}}
\put(2,4.0){\line(1,0){0.5}}
\put(2,4.5){\line(1,0){0.5}}
\put(2,5.0){\line(1,0){0.5}}
\put(2,5.5){\line(1,0){0.5}}
\put(2,6.0){\line(1,0){0.5}}
\put(2,6.5){\line(1,0){0.5}}
\put(2,0.5){\line(0,1){6}}
\put(2.5,0.5){\line(0,1){6}}
     
\put(1.3,0.0){\makebox(0,0){(a) $S_{N_c(N_f+Q)}$}}

\put(6.75,7.0){\vector(1,0){0.75}}
\put(5.25,7.0){\vector(-1,0){0.75}}

\put(6.0,7.0){\makebox(0,0){$N_f+Q$}}

\put(4.0,5.0){\vector(0,1){1.5}}
\put(4.0,4.5){\vector(0,-1){1.5}}

\put(4.0,4.75){\makebox(0,0){$N_c$}}

\put(4.5,3.0){\line(1,0){3}}
\put(4.5,3.5){\line(1,0){3}}
\put(4.5,4.0){\line(1,0){3}}
\put(4.5,4.5){\line(1,0){3}}
\put(4.5,5.0){\line(1,0){3}}
\put(4.5,5.5){\line(1,0){3}}
\put(4.5,6.0){\line(1,0){3}}
\put(4.5,6.5){\line(1,0){3}}
\put(4.5,3.0){\line(0,1){3.5}}
\put(5.0,3.0){\line(0,1){3.5}}
\put(5.5,3.0){\line(0,1){3.5}}
\put(6.0,3.0){\line(0,1){3.5}}
\put(6.5,3.0){\line(0,1){3.5}}
\put(7.0,3.0){\line(0,1){3.5}}
\put(7.5,3.0){\line(0,1){3.5}}

\put(6.0,0.0){\makebox(0,0){(b) SU($N_c$)}}

\put(11.5,7.0){\vector(1,0){1.5}}
\put(11.0,7.0){\vector(-1,0){1.5}}

\put(11.25,7.0){\makebox(0,0){$N_c$}}

\put(9,5.25){\vector(0,1){1.25}}
\put(9,4.75){\vector(0,-1){1.25}}

\put(8.7,5.0){\makebox(0,0){$N_f+Q$}}

\put(9.5,3.5){\line(1,0){3.5}}
\put(9.5,4.0){\line(1,0){3.5}}
\put(9.5,4.5){\line(1,0){3.5}}
\put(9.5,5.0){\line(1,0){3.5}}
\put(9.5,5.5){\line(1,0){3.5}}
\put(9.5,6.0){\line(1,0){3.5}}
\put(9.5,6.5){\line(1,0){3.5}}
\put(9.5,3.5){\line(0,1){3}}
\put(10.0,3.5){\line(0,1){3}}
\put(10.5,3.5){\line(0,1){3}}
\put(11.0,3.5){\line(0,1){3}}
\put(11.5,3.5){\line(0,1){3}}
\put(12.0,3.5){\line(0,1){3}}
\put(12.5,3.5){\line(0,1){3}}
\put(13.0,3.5){\line(0,1){3}}

\put(11.25,0.0){\makebox(0,0){(c) U($2N_f$)}}

\end{picture}
\vspace*{2mm}
\caption{Irreducible representations defined by the different $Q$ sectors.}
\label{fig:young-graph}
\end{center}
\end{figure}

From Eq.~\eqref{eq:color-neutral}, we see that the color-neutral
sector splits into subsectors labeled by $Q$, with fermion occupation
number $N_c(N_f +Q)$, respectively.  Hence, the Young diagram of a
representation of a group acting irreducibly in such a subsector of
Fock space necessarily has $N_c(N_f+Q)$ boxes.  Since the fermionic
operators obey anticommutation relations, permutations of the fermions
furnish a totally antisymmetric representation of the permutation
group $S_{N_c(N_f+Q)}$, with the Young diagram shown in
Fig.~\ref{fig:young-graph}(a).

The action of the group SU($N_c$) on a state in the subsector labeled
by $Q$ leaves that state invariant.  Therefore, the representation of
SU($N_c$) in this sector is the trivial one.  Again, the corresponding
Young diagram is made up of $N_c(N_f+Q)$ boxes.  It is shown in
Fig.~\ref{fig:young-graph}(b).  Viewed as a representation of
$S_{N_c(N_f+Q)}$, it is totally symmetric under permutations of flavor
indices and totally antisymmetric under permutations of color indices.

We now seek to determine the representation of U($2N_f$) acting in the
subsector labeled by $Q$.  We will conclude below that this
representation is irreducible, but for the moment we leave open the
possibility that it may be reducible, in which case we decompose it
into irreducible parts labeled by their corresponding Young diagrams.
These Young diagrams also define representations of $S_{N_c(N_f+Q)}$.
The subsectors labeled by $Q$ are invariant subspaces under the action
of SU($N_c$) $\times$ U($2N_f$) (recall that the generators of these
two groups commute).  Therefore, we obtain a representation of
$S_{N_c(N_f+Q)}$ by taking the direct product of the representation of
$S_{N_c(N_f+Q)}$ shown in Fig.~\ref{fig:young-graph}(b) and the direct
sum of the representations of $S_{N_c(N_f+Q)}$ given by the Young
diagrams corresponding to the irreducible representations of U($2N_f$)
contained in the decomposition mentioned above.  The representation of
$S_{N_c(N_f+Q)}$ thus obtained must contain the totally antisymmetric
representation given by the Young diagram in
Fig.~\ref{fig:young-graph}(a).

We now use the fact that the totally antisymmetric representation of
the permutation group is contained only in the direct product of a
given irreducible representation with its associate, which has the
transposed Young diagram.  Therefore, we conclude that the direct sum
representation of U($2N_f$) mentioned above must contain the
irreducible representation shown in Fig.~\ref{fig:young-graph}(c).  In
fact, it cannot contain any other irreducible representation, since
the direct product of the representation given by
Fig.~\ref{fig:young-graph}(b) with such a representation does not lead
to a totally antisymmetric representation of $S_{N_c(N_f+Q)}$.  Also,
the representation of Fig.~\ref{fig:young-graph}(c) cannot occur more
than once, since the direct product with Fig.~\ref{fig:young-graph}(b)
would then lead to more than one totally antisymmetric representations
of $S_{N_c(N_f+Q)}$.  This possibility can be excluded because
$S_{N_c(N_f+Q)}$ acts irreducibly in the subsector labeled by $Q$.
Thus, Fig.~\ref{fig:young-graph}(c) defines an irreducible
representation of U($2N_f$) in this subsector.  When referring to this
representation, we shall simply denote it by $Q$.

The corresponding highest-weight vector is obtained by acting on the
Fock vacuum with a product of $N_c(N_f+Q)$ creation operators
\cite{Read:1989jy},
\begin{equation}
\label{eq:highest_weight}
\ket{\psi_Q}\equiv (\bar{c}_1^1 \bar{c}_2^1\cdots \bar{c}_{N_f+Q}^1)
(\bar{c}_1^2 \bar{c}_2^2\cdots \bar{c}_{N_f+Q}^2)
\cdots
(\bar{c}_1^{N_c} \bar{c}_2^{N_c}\cdots \bar{c}_{N_f+Q}^{N_c})\ket{0}\:.
\end{equation}

The action of the group U($2N_f$) on the Fock space is now defined by
the following mapping of group elements $g\in\text{U}(2N_f)$ to
operators $T_g$,
\begin{equation}
T_g = \exp\left( \bar{c}_A^i (\log g)_{AB}c_B^i \right)\:.
\end{equation}
Zirnbauer showed in Ref.~\cite{Zirnbauer:1996} that this is indeed a
well-defined map, and furthermore, that it is a homomorphism, i.e.\ 
$T_{g_1}T_{g_2}=T_{g_1 g_2}$.  Thus, the mapping $g\mapsto T_g$
defines a (reducible) representation of U($2N_f$).  We choose the
color-neutral sector to be the carrier space for this
representation. It decomposes 
into invariant subspaces, the carrier spaces of the (irreducible)
representations $Q$.  The $\bar{c}_A^i$ obey the
important transformation property $T_g \bar{c}_A^i
T_g^{-1}=\bar{c}_B^i g_{BA}$, and similarly we have $T_g c_{A}^i
T_g^{-1}=g^{\dag}_{AB}c_B^i$.

Next, we set up a system of generalized coherent states for each
representation $Q$ defined above. A comprehensive introduction to
generalized coherent states is given in Ref.~\cite{Perelomov:1986}.
They form an overcomplete set of states, which have a number of nice
properties. Most importantly for our purposes, they allow a resolution
of the identity operator.

For a general Lie group $G$ and an irreducible unitary representation
$T_g$ of $G$, such a set of states is obtained as follows. Take any
state in the carrier space of $T_g$, say $\ket{\psi_T}$, and act with
all elements $T_g$ on this state, resulting in the set $\{
T_g\ket{\psi_T}\}$. In general, not all of the resulting states will
be distinct. Let $H$ be the maximal subgroup of $G$ such that for all
$h\in H, \; T_h\ket{\psi_T}\propto \ket{\psi_T}$.  The subgroup $H$ is
called the isotropy subgroup of $\ket{\psi_T}$, and the set of
coherent states may be parameterized without overcounting by the
elements of the coset space $G/H$. A particularly convenient state to
pick as the starting vector is the highest- (or lowest-) weight vector
of the representation $T_g$. In this case, the corresponding isotropy
subgroup will contain the Cartan subgroup of $G$.  (Note that it is
also possible to work with a subgroup $H$ that is not maximal.  In
this case a parameterization by the elements of $G/H$ would not
eliminate all the redundant states. This is of no concern as long as
for integrals over $G/H$ the measure of integration is suitably normalized.)

Now consider the representation $Q$ of U($2N_f$). The highest-weight
vector is given in Eq.~\eqref{eq:highest_weight}, and hence the
corresponding set of coherent states is $\{ T_g\ket{\psi_Q}\}$.  The
resolution of the identity takes the form \cite{Perelomov:1986}
\begin{equation}
\1_Q = C_Q
\int_{\text{U}(2N_f)}d\mu(g)T_g\ket{\psi_Q}\bra{\psi_Q}T_g^{-1}\:,
\end{equation}
where $d\mu(g)$ is the invariant Haar measure for the group U($2N_f$)
and $C_Q$ is a normalization constant which will be
determined explicitly in Eq.~\eqref{eq:proj_const}.
We normalize the measure so that the group volume is unity. 
The proof that $\1_Q$ must be proportional to the identity uses
the fact that it commutes with all group elements, by virtue
of the invariance of the measure, and an invocation of Schur's lemma.
Note that $\1_Q$ is actually a projector onto the part of Fock space
that forms the carrier space of $Q$ since it not only is the identity
operator in this sector but also annihilates all states outside this
sector. (This is easy to see, because it clearly annihilates states which
are not color neutral, as well as states with a different occupation
number, i.e.\ a different value of $Q$.)  The projection $P$ onto the
SU($N_c$) invariant sector of Fock space can therefore be written as a
sum over $Q$ of the $\1_Q$'s,
\begin{equation}
\label{eq:sumproj}
P=\sum_{Q=-N_f}^{N_f}\1_Q\:.
\end{equation}

In the next section, we show 
how to remove the redundancy in the
coherent states and how to parameterize them.

\subsection{Parameterization of coherent states in terms of coset spaces}
\label{cosets}
As mentioned in the previous section, there is some redundancy in the
set $\{T_g\ket{\psi_Q}\}$.  Let us first focus on the case of
$Q=0$ and determine the coset space U$(2N_f)/H$ needed to label
the corresponding coherent states. The highest-weight vector of the
representation $Q=0$ is
\begin{equation}
\ket{\psi_0}\equiv (\bar{c}_1^1 \bar{c}_2^1\cdots \bar{c}_{N_f}^1)
(\bar{c}_1^2 \bar{c}_2^2\cdots \bar{c}_{N_f}^2)
\cdots
(\bar{c}_1^{N_c} \bar{c}_2^{N_c}\cdots \bar{c}_{N_f}^{N_c})\ket{0}\:,
\end{equation}
and thus, this state has the property that $\bar{c}_a^i\ket{\psi_0}
=d_b^j\ket{\psi_0}=0$. It immediately follows that for elements of
U($2N_f$) of the form $h=\diag(h_{+},h_{-})$, where $h_{\pm}$ are
unitary $N_f\times N_f$ matrices, the corresponding Fock operators
$T_h$ leave the state $\ket{\psi_0}$ unchanged,
\begin{align}
T_h\ket{\psi_0}&=\exp\left(\bar{c}_a^i(\log h_{+})_{ab}c_b^i
+\bar{d}_a^i(\log h_{-})_{ab}d_b^i\right)\ket{\psi_0} \notag \\
&=\exp\left(N_c \Tr(\log(h_{+})\right)\ket{\psi_0} =
\ket{\psi_0}\det(h_{+})^{N_c}. 
\end{align}
Hence, we can parameterize the coherent states by elements of $G/H$,
where $G=\text{U}(2N_f)$ and $H=\text{U}(N_f)\times \text{U}(N_f)$. To
do so, we follow Ref.~\cite{Zirnbauer:1996} and define $\pi:G\to G/H$
to be the canonical projection which assigns to each group element
$g\in G$ the corresponding equivalence class $gH$.  Choosing a
representative group element $s(\pi(g))$ from each such equivalence
class, we decompose an arbitrary group element $g$ into a product
$g=s(\pi(g))h(g)$, where $s(\pi(g))$ takes values in $G$ and $h(g)$
takes values in $H$.  The map $s:G/H\to G$ is called a local section
of the bundle $\pi:G\to G/H$.

In order to arrive at an expression for $s(\pi(g))$, write the 
$2N_f\times 2N_f$ dimensional matrix $g$ in terms of $N_f\times N_f$
dimensional blocks,
\begin{equation}
g=\mat{a}{b}{c}{d}\:.
\end{equation}
Multiplying $g$ by $h=\diag(h_{+},h_{-})$ on the right leaves the
products $Z=ca^{-1}$ and $\tilde{Z}=bd^{-1}$ invariant. From the
unitarity condition for $g$, it is also immediate that
$\tilde{Z}=-Z^{\dag}$.  A particular $s(\pi(g))$ is then given by
\begin{align}
\label{eq:s}
s(\pi(g))\equiv s(Z,Z^{\dag})&=
\mat{(1+Z^{\dag}Z)^{-1/2}}{-Z^{\dag}(1+ZZ^{\dag})^{-1/2}}
{Z(1+Z^{\dag}Z)^{-1/2}}{(1+ZZ^{\dag})^{-1/2}} \\
&=\mat{1}{0}{Z}{1}\mat{(1+Z^{\dag}Z)^{-1/2}}{0}{0}{(1+ZZ^{\dag})^{+1/2}}
\mat{1}{-Z^{\dag}}{0}{1}\:. \notag
\end{align}
(Note that this parameterization differs slightly from Zirnbauer's
notation, because we have chosen the state $\ket{\psi_0}$ so
that it is annihilated by $\bar{c}_a^i$ and $d_a^i$, which reverses
the roles of $c_a^i$ and $d_a^i$ with respect to Ref.~\cite{Zirnbauer:1996}.)
We can thus label the coherent states for $Q=0$ by $Z$, 
and obtain
\begin{align}
\ket{Z}&\equiv T_{s(Z,Z^{\dag})}\ket{\psi_0} \notag \\
&=\exp\left(\bar{d}_a^i Z_{ab}c_b^i\right)\exp\left(-\frac{1}{2}\bar{c}_a^i
\log(1+Z^{\dag}Z)_{ab}c_b^i \right. \notag \\
&\hspace{4.2cm}\left.+\frac{1}{2}\bar{d}_a^i\log(1+ZZ^{\dag})_{ab}
d_b^i\right)\exp\left(-\bar{c}_a^i Z^{\dag}_{ab}d_b^i\right)
\ket{\psi_0}\notag \\ 
&=\exp(\bar{d}_a^i Z_{ab} c_b^i)\ket{\psi_0}\det(1+Z^{\dag}Z)^{-N_c/2}\:.
\end{align}
Similarly, we have
\begin{equation}
\bra{Z}\equiv \bra{\psi_0}T_{s(Z,Z^{\dag})}^{-1}
=\det(1+ZZ^{\dag})^{-N_c/2}\bra{\psi_0}\exp\left( \bar{c}_a^i
Z_{ab}^{\dag}d_b^i\right)\:.
\end{equation}
Regarding integration over the coset space U$(2N_f)/$U$(N_f)\times
$U$(N_f)$, a U$(2N_f)$ invariant measure exists so that 
\begin{equation}
\int_{\text{U}(2N_f)}d\mu(g)f(g)=\int_{\text{U}(2N_f)/H}D(Z,Z^{\dag})
\int_{H}d\mu(h)f(s(\pi(g))h(g))\:.
\end{equation}
In fact, this measure is given
by (cf.\ \cite{Zirnbauer:1996,Budczies:2000qs}) 
\newcounter{saveeqnum}
\setcounter{saveeqnum}{\value{equation}}
\renewcommand{\theequation}{\ref{eq:measure}}
\begin{equation}
  D(Z,Z^{\dag})=C\:\frac{dZdZ^\dagger}{\det(1+ZZ^{\dag})^{2N_f}}
  \quad\text{with}\quad
  dZdZ^\dagger=\prod_{a,b=1}^{N_f}d\re Z_{ab}\;d\im Z_{ab}\:.
\end{equation}
\setcounter{equation}{\value{saveeqnum}}%
\renewcommand{\theequation}{\arabic{equation}}%
As mentioned in Sec.~\ref{sec:result}, the constant $C$ is chosen such
that $\int D(Z,Z^\dagger)=1$.  An explicit expression for $C$ is
\begin{equation}
  \label{eq:normNf}
  C=\frac1{\pi^{N_f^2}}\prod_{n=0}^{N_f-1}\frac{(N_f+n)!}{n!}\:,
\end{equation}
as computed, among many other results, in a very interesting article
\cite{BNSZ} which appeared after the original submission of the
present paper.

The resolution of unity in the $Q=0$ sector thus becomes
\begin{align}
\1_0 &= C_0\int_{\text{U}(2N_f)}d\mu(g)T_g\ket{\psi_0}\bra{\psi_0}
T_g^{-1} \notag \\ 
&=C_0\int_{\text{U}(2N_f)/H}D(Z,Z^{\dag})T_{s(Z,Z^{\dag})}\ket{\psi_0}
\bra{\psi_0}T_{s(Z,Z^{\dag})}^{-1} \notag\\
&=C_0\int D(Z,Z^{\dag})\ket{Z}\bra{Z}\:,
\end{align}
where we have set $\vol(H)=1$.  As mentioned in Sec.~\ref{sec:result},
our choice of normalization implies that
\begin{equation}
\label{eq:norm}
C_0\int \frac{D(Z,Z^{\dag})}{\det(1+ZZ^{\dag})^{N_c}}=1\:,
\end{equation}
consistent with $\vol(\text{SU}(N_c))=1$.  The constant $C_0$ is
given in Eq.~\eqref{eq:norm2}.

This essentially completes the discussion of the $Q=0$ sector which
was necessary to derive the color-flavor transformation for U($N_c$).
For the gauge group SU($N_c$) we now have to consider the
nonzero values of $Q$ in Eq.~\eqref{eq:sumproj}.  The isotropy
subgroup for the highest-weight state belonging to representations
with $Q\neq 0$ is not U$(N_f)\times$U$(N_f)$ but rather
$H_Q=$U$(N_f+Q)\times$U$(N_f-Q)$.  Parameterizing the coherent states
in terms of the coset spaces U$(2N_f)/H_Q$ would result in a sum over
$Q$ of integrals with different parameterizations and different
integration domains for each $Q$.  This would lead to a correct result
whose usefulness, however, is rather questionable. Instead, we
seek to express the projectors $\1_Q$ in terms of the 
variables $Z$, $Z^{\dag}$ just derived and
integrate over U$(2N_f)/H_{Q=0}$.  This can be achieved by
relating the highest-weight vector for $Q\neq 0$ to $\ket{\psi_0}$
by the action of an appropriate combination of creation or annihilation 
operators on $\ket{\psi_0}$. Consider for instance the case of $Q>0$,
\begin{align}
  \1_Q&=C_Q\int_{\text{U}(2N_f)}d\mu(g)T_g\ket{\psi_Q}\bra{\psi_Q}
      T_g^{-1} \notag \\ 
      &=C_Q\int_{\text{U}(2N_f)}d\mu(g)T_g\left(\bar{d}_Q^1\cdots
      \bar{d}_Q^{N_c}\right) 
        \cdots\left(\bar{d}_1^1\cdots \bar{d}_1^{N_c}\right)\ket{\psi_0}
        \notag \\
       &\hspace{5cm}\bra{\psi_0}\left(d_1^{N_c}\cdots d_1^1\right)
        \cdots\left(d_Q^{N_c}\cdots d_Q^1\right)T_g^{-1}\:.
  \label{eq:project-q}
\end{align}
We can now insert $T_g^{-1}T_g$ between each pair of $\bar{d}$'s and
each pair of $d$'s. We also insert such a term between
$\bar{d}_1^{N_c}$ and $\ket{\psi_0}$, and between $\bra{\psi_0}$ and
$d_1^{N_c}$. Now, make use of the fact that $T_g \bar{d}^i_a
T_g^{-1}=\bar{c}^i_{A^i_{a}}g_{A^i_{a},N_f+a}$. Recall that
$A^i_{a}$ is a label which runs over all values from 1 to $2N_f$.
Similarly, $T_g d_a^i
T_g^{-1}=g^{\dag}_{N_f+a,B^i_{a}}c^i_{B^i_{a}}$.  Next, we
perform a coset decomposition of the group elements $g$,
$g=s(Z,Z^{\dag})h(g)$.  Note that
$T_g\ket{\psi_0}\bra{\psi_0}T_g^{-1}=\ket{Z}\bra{Z}$. Thus, we arrive
at the following expression for the projector $\1_Q$ in the case of
$Q>0$,
\begin{align}
  \1_Q&=C_Q
       \int D(Z,Z^{\dag})\bigg[\int_{\text{U}(N_f)}\!\!\!d\mu(h_{-})
 \widetilde{\Gamma}^{(a^1_{Q}\cdots a^{N_c}_{Q})\cdots(a^{1}_{1}\cdots
  a^{N_c}_{1})}_{(Q\cdots Q)\cdots(1 \cdots 1)}
 \left(\widetilde{\Gamma}^{(b^1_{Q}\cdots b^{N_c}_{Q})\cdots(b^{1}_{1}
  \cdots b^{N_c}_{1})}_{(Q \cdots Q)\cdots(1\cdots 1)}\right)^{*} 
  \bigg]\notag \\ 
  &\qquad\qquad\qquad\quad(s_{A^1_{Q},N_f+a^1_{Q}}\cdots
    s_{A^{N_c}_{Q},N_f+a^{N_c}_{Q}})\cdots
    (s_{A^{1}_{1},N_f+a^{1}_{1}}\cdots s_{A^{N_c}_{1},N_f+a^{N_c}_{1}})
    \notag \\
    &\qquad\qquad\qquad\quad(s^{\dag}_{N_f+b^{N_c}_{Q},B^{N_c}_{Q}}\cdots 
      s^{\dag}_{N_f+b^{1}_{Q},B^1_{Q}})\cdots
     (s^{\dag}_{N_f+b^{N_c}_{1},B^{N_c}_{1}} \cdots 
      s^{\dag}_{N_f+b^{1}_{1},B^1_{1}} )  \notag \\
    &\quad\left(\bar{c}^1_{A^1_{Q}}\cdots\bar{c}^{N_c}_{A^{N_c}_{Q}}\right)
    \cdots
    \left(\bar{c}^1_{A^1_{1}}\cdots\bar{c}^{N_c}_{A^{N_c}_{1}}\right)
    \!\ket{Z} 
    \bra{Z}\!\left(c^{N_c}_{B^{N_c}_{1}}
    \cdots c^{1}_{B^1_{1}}\right)\cdots
    \left(c^{N_c}_{B^{N_c}_{Q}}\cdots c^{1}_{B^1_{Q}}\right),
\label{eq:q_projector}
\end{align}
where $\widetilde{\Gamma}=\widetilde{\Gamma}(h_-)$ denotes the
representation matrix of the tensor product of $N_c\cdot Q$
fundamental representations of U($N_f$) (cf.\ 
Fig.~\ref{fig:u-n-young-graph}),
\begin{equation}
  \widetilde{\Gamma}^{(a^1_{Q}\cdots a^{N_c}_{Q})\cdots(a^{1}_{1}\cdots
  a^{N_c}_{1})}_{(b^1_{Q}\cdots b^{N_c}_{Q})\cdots(b^{1}_{1}\cdots
  b^{N_c}_{1})}\equiv \bigl((h_{-})_{a^1_{Q},b^1_{Q}}\cdots 
  (h_{-})_{a^{N_c}_{Q},b^{N_c}_{Q}}\bigr) 
  \cdots \bigl((h_{-})_{a^1_{1},b^1_{1}}\cdots 
  (h_{-})_{a^{N_c}_{1},b^{N_c}_{1}}\bigr)\:.
\label{eq:gamma_tensor}
\end{equation}
The case of $Q<0$ is very similar, the analog of Eq.~\eqref{eq:project-q}
being
\begin{multline}
  \1_Q=C_Q
        \int_{\text{U}(2N_f)}d\mu(g)T_g\left(c_{N_f+Q+1}^1\cdots
        c_{N_f+Q+1}^{N_c}\right) 
        \cdots\left(c_{N_f}^1\cdots c_{N_f}^{N_c}\right)\ket{\psi_0}\\
       \bra{\psi_0}\left(\bar{c}_{N_f}^{N_c}\cdots 
        \bar{c}_{N_f}^{1}\right)
        \cdots\left(\bar{c}_{N_f+Q+1}^{N_c}\cdots \bar{c}_{N_f+Q+1}^1\right)
        T_g^{-1}\:.
\end{multline}

\subsection{Fermi coherent states}
\label{fermi}
We shall also need Fermi coherent states. These are defined as
\begin{equation}
  \exp\left(\bar{\varphi}_a^i c_a^i +\bar{d}_a^i\psi_a^i\right)\ket{\psi_0}\:,
\end{equation}
and they span the entire Fock space. In the same way that averaging
over rotations of a vector in $\mathbb{R}^3$ around the $z$-axis
projects out the $z$-component of that vector, i.e.\ the part that is
invariant under such a rotation, we can project out the color-neutral
part of a Fermi coherent state by averaging over SU($N_c$) rotations,
\begin{equation}
  \label{eq:P2}
  P\exp\left(\bar{\varphi}_a^i c_a^i +\bar{d}_a^i\psi_a^i\right)\ket{\psi_0}
 = \int_{\text{SU}(N_c)}dU \exp\left(\bar{\varphi}_a^i U^{\dagger ij} c_a^j 
+\bar{d}_a^i U^{ij}\psi_a^j\right)\ket{\psi_0}.
\end{equation}
The identification of the two projection operators in
Eqs.~\eqref{eq:sumproj} and \eqref{eq:P2} now enables us to complete
the proof of the color-flavor transformation.

\section{Proof of the result}
\label{sec:proof}
Starting from the left-hand side of Eq.~\eqref{eq:col_flav}, we obtain
\begin{align}
  I &=\int_{\text{SU}(N_c)}dU \exp\left( \bar{\psi}_a^i U^{ij}\psi_a^j
+\bar{\varphi}_a^i U^{\dagger ij}\varphi_a^j \right) \notag \\
&=\int_{\text{SU}(N_c)}dU\braopket{\psi_0} {\exp\left(\bar{\psi}_a^i
    d_a^i-\bar{c}_a^i 
\varphi_a^i \right)\exp\left(\bar{d}_a^iU^{ij}\psi_a^j+\bar{\varphi}_a^i 
U^{\dagger ij} c_a^j\right)}{\psi_0} \notag \\
&=\braopket{\psi_0}{\exp\left(\bar{\psi}_a^i
    d_a^i-\bar{c}_a^i\varphi_a^i \right) 
P\exp\left(\bar{d}_a^i \psi_a^i+\bar{\varphi}_a^i c_a^i\right)}{\psi_0} 
\notag \\
&=\sum_{Q=-N_f}^{N_f}
\braopket{\psi_0}{\exp\left(\bar{\psi}_a^i d_a^i-\bar{c}_a^i\varphi_a^i \right)
\1_Q\exp\left(\bar{d}_a^i \psi_a^i+\bar{\varphi}_a^i
  c_a^i\right)}{\psi_0} \:.
\label{eq:short_proof}
\end{align}
The second line is obtained from the first one using the
anticommutation relations and the action of the creation and
annihilation operators on the state $\ket{\psi_0}$.  To obtain the
third line, we have used Eq.~\eqref{eq:P2}.  The implementation of the
projection operator onto the color-neutral sector given by
Eq.~\eqref{eq:sumproj} then yields the fourth line.

\begin{figure}[tb]
\begin{center}
\unitlength1cm
\begin{picture}(7.1,4)
\put(0.0,2.0){\makebox(0,0){$\widetilde{\Gamma}\::$}}
\put(0.5,1.1){$\begin{array}{c}
\underbrace{
\begin{picture}(3.5,0.5)(0,-0.15)
\put(0.0,0.0){\line(1,0){0.5}}
\put(0.0,0.5){\line(1,0){0.5}}
\put(0.0,0.0){\line(0,1){0.5}}
\put(0.5,0.0){\line(0,1){0.5}}
\put(0.75,0.25){\makebox(0,0){$\otimes$}}
\put(1.0,0.0){\line(1,0){0.5}}
\put(1.0,0.5){\line(1,0){0.5}}
\put(1.0,0.0){\line(0,1){0.5}}
\put(1.5,0.0){\line(0,1){0.5}}
\put(1.75,0.25){\makebox(0,0){$\otimes$}}
\put(2.25,0.25){\makebox(0,0){$\cdots$}}
\put(2.75,0.25){\makebox(0,0){$\otimes$}}
\put(3.0,0.0){\line(1,0){0.5}}
\put(3.0,0.5){\line(1,0){0.5}}
\put(3.0,0.0){\line(0,1){0.5}}
\put(3.5,0.0){\line(0,1){0.5}}
\end{picture}}\\ 
N_c\cdot Q 
\end{array}$}
\put(4.5,1.8){\vector(1,0){1.25}}
\put(6.5,1.8){\makebox(0,0){$\Gamma\::$}}
\end{picture}
\begin{picture}(5,4)
\put(3.0,4.0){\vector(1,0){2}}
\put(2.5,4.0){\vector(-1,0){2}}

\put(2.75,4.0){\makebox(0,0){$N_c$}}

\put(0.0,2.0){\vector(0,1){1.5}}
\put(0.0,1.5){\vector(0,-1){1.5}}

\put(0.0,1.75){\makebox(0,0){$Q$}}

\put(0.5,0.0){\line(1,0){4.5}}
\put(0.5,0.5){\line(1,0){4.5}}
\put(0.5,1.0){\line(1,0){4.5}}
\put(0.5,1.5){\line(1,0){4.5}}
\put(0.5,2.0){\line(1,0){4.5}}
\put(0.5,2.5){\line(1,0){4.5}}
\put(0.5,3.0){\line(1,0){4.5}}
\put(0.5,3.5){\line(1,0){4.5}}
\put(0.5,0.0){\line(0,1){3.5}}
\put(1.0,0.0){\line(0,1){3.5}}
\put(1.5,0.0){\line(0,1){3.5}}
\put(2.0,0.0){\line(0,1){3.5}}
\put(2.5,0.0){\line(0,1){3.5}}
\put(3.0,0.0){\line(0,1){3.5}}
\put(3.5,0.0){\line(0,1){3.5}}
\put(4.0,0.0){\line(0,1){3.5}}
\put(4.5,0.0){\line(0,1){3.5}}
\put(5.0,0.0){\line(0,1){3.5}}
\end{picture}
\caption{The (reducible) product of fundamental representations of
  U($N_f$) contains an irreducible representation of U($N_f$) with
  symmetric color indices and antisymmetric flavor indices.  Here
  $Q>0$.}
\label{fig:u-n-young-graph}
\end{center}
\end{figure}
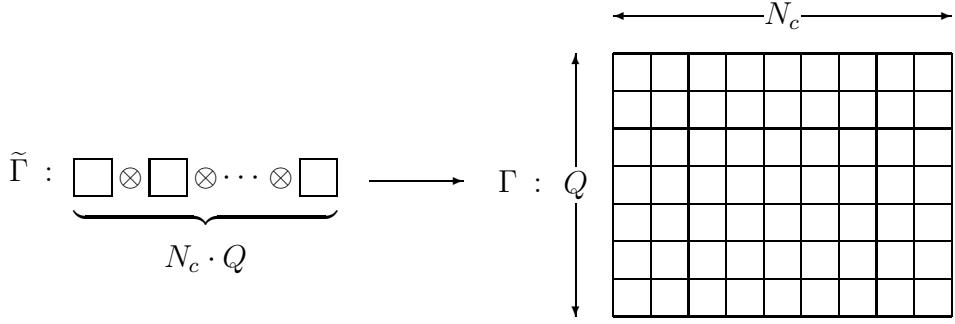

The next step is to evaluate each term in the sum over $Q$. As before,
consider the case of $Q>0$. From
Eqs.~\eqref{eq:q_projector} and \eqref{eq:short_proof}, 
we need to consider matrix elements of the
form
\begin{align}
&\braopket{\psi_0}{\exp\left(\bar{\psi}_a^i
    d_a^i-\bar{c}_a^i\varphi_a^i \right) 
    \left(\bar{c}^1_{A^1_{Q}}\cdots\bar{c}^{N_c}_{A^{N_c}_{Q}}\right)\cdots
    \left(\bar{c}^1_{A^1_{1}}\cdots\bar{c}^{N_c}_{A^{N_c}_{1}}\right)
    \exp\left(\bar{d}_a^i Z_{ab}c_b^i \right)}{\psi_0} \notag\\
  &\qquad\qquad=\exp\left(\bar{\psi}_a^i Z_{ab} \varphi_b^i\right) 
  \left(\bar{\Psi}_{A^1_{Q}}^1\cdots
    \bar{\Psi}_{A^{N_c}_{Q}}^{N_c}\right)    
  \cdots \left(\bar{\Psi}_{A^1_{1}}^1\cdots\bar{\Psi}_{A^{N_c}_{1}}^{N_c} 
  \right)\:,
\end{align}
where we have again used the properties of the creation and
annihilation operators.  We have also introduced the shorthand
notation
\begin{equation}
  \bar{\Psi}^i = \left( -\bar{\psi}^i Z\;,\; \bar{\psi}^i \right)
\end{equation}
for $\bar{\Psi}^i_a=-\bar\psi^i_bZ_{ba}$ and
$\bar{\Psi}^i_{N_f+a}=\bar\psi^i_{a}$.
Similarly, we find that
\begin{align}
 & \braopket{\psi_0}{\exp\left(\bar{c}^i_a Z^{\dag}_{ab}d^i_b\right)
  \left(c^{N_c}_{B^{N_c}_{1}}
    \cdots c^{1}_{B^1_{1}}\right)\cdots
    \left(c^{N_c}_{B^{N_c}_{Q}}\cdots c^{1}_{B^1_{Q}}\right) 
\exp\left(\bar{d}_a^i \psi_a^i+\bar\varphi_a^i c_a^i \right)}{\psi_0} \notag\\
&\qquad\qquad=\exp\left( -\bar\varphi_a^i Z^{\dag}_{ab}\psi_b^i\right)
\left(\Psi_{B^{N_c}_{1}}^{N_c}\cdots \Psi_{B^{1}_{1}}^1 \right)\cdots 
\left(\Psi_{B^{N_c}_{Q}}^{N_c}\cdots \Psi_{B^{1}_{Q}}^1 \right)\:,
\end{align}
where 
\begin{equation}
  \Psi^i = \mvec{-Z^{\dag}\psi^i}{\psi^i}\:.
\end{equation}
We also need the matrix-vector products
\begin{equation}
  \label{eq:s_psi}
  \bar{\Psi}^i s(Z,Z^{\dag})=(0,\overline{\widehat{\psi}}^i)
  \qquad\mbox{and}\qquad
  s(Z,Z^{\dag})^{\dag}\Psi^i=\mvec{0}{\widehat{\psi}^i}\:,
\end{equation}
where $\overline{\widehat{\psi}}^i=\bar{\psi}^i(1+ZZ^{\dag})^{1/2}$
and $\widehat{\psi}^i=(1+ZZ^{\dag})^{1/2}\psi^i$.  Putting everything
together and using Eq.~\eqref{eq:q_projector}, we find that
\begin{equation}
  I = C_0\int \frac{D(Z,Z^{\dag})}{\det(1+ZZ^\dagger)^{N_c}}
  \exp\left(\bar{\psi}_a^i Z_{ab}\varphi_b^i 
-\bar{\varphi}_a^i Z_{ab}^{\dag}\psi_b^i \right)
\sum_{Q=-N_f}^{N_f} \chi_Q\:,
\end{equation}
where for $Q>0$, we have
\begin{align}
  \chi_Q=&\frac{C_Q}{C_0}
  \left[\int_{\text{U}(N_f)}d\mu(h_{-})
  \widetilde{\Gamma}^{(a^1_{Q}\cdots a^{N_c}_{Q})\cdots(a^{1}_{1}\cdots
  a^{N_c}_{1})}_{(Q\cdots Q)\cdots(1 \cdots 1)}
  \left(\widetilde{\Gamma}^{(b^1_{Q}\cdots b^{N_c}_{Q})\cdots(b^{1}_{1}
  \cdots b^{N_c}_{1})}_{(Q \cdots Q)\cdots(1\cdots 1)}\right)^{*} 
  \right] \notag \\
  &\left(\oh{\psi}_{a^1_{Q}}^1 \cdots \oh{\psi}_{a^{N_c}_{Q}}^{N_c} \right) 
  \cdots \left(\oh{\psi}_{a^1_{1}}^1\cdots  \oh{\psi}_{a^{N_c}_{1}}^{N_c} 
  \right) 
  \left(\wh{\psi}_{b^{N_c}_{1}}^{N_c}\cdots \wh{\psi}_{b^1_{1}}^1 \right)
  \cdots\left(\wh{\psi}_{b^{N_c}_{Q}}^{N_c}\cdots \wh{\psi}_{b^1_{Q}}^{1}
  \right)\:.
\end{align}
From the definition of $\widetilde{\Gamma}$ in
Eq.~\eqref{eq:gamma_tensor} it is clear that the expression in square
brackets is totally symmetric under the exchange of $a^i_{a}$ with
$a^{i'}_{a}$, and of $b^i_{a}$ with $b^{i'}_{a}$. Furthermore, the
product of the $\oh{\psi}$'s is totally antisymmetric under exchange
of $a^i_{a}$ with $a^i_{a'}$, since the $\oh{\psi}$'s are Grassmann
variables. Similarly, the product of the $\wh{\psi}$'s is
antisymmetric with respect to the interchange of $b^i_{a}$ with
$b^i_{a'}$. In each term, we now pick out only those pieces of
$\widetilde{\Gamma}$ that have the correct symmetry properties, i.e.\ 
symmetric in color and antisymmetric in flavor.  To this end, we
decompose the (reducible) product of fundamental representations of
U($N_f$), represented by $\widetilde{\Gamma}$, into irreducible
representations.  Of these, only the one with the correct symmetry
properties, shown in Fig.~\ref{fig:u-n-young-graph}, survives the
contraction of indices.  We can thus replace $\widetilde{\Gamma}$ by
the representation matrix corresponding to this irreducible
representation, which we denote by $\Gamma$, and obtain
\begin{align}
  \chi_Q=\frac{C_Q}{C_0} &
  \left\{ \frac{1}{(N_c!)^Q}\sum_{\sigma_{1}\cdots\sigma_{Q}}
  \left(\oh{\psi}_{\sigma_{Q}(a^1_{Q})}^1 \cdots 
  \oh{\psi}_{\sigma_{Q}(a^{N_c}_{Q})}^{N_c} \right) 
  \cdots \left(\oh{\psi}_{\sigma_{1}(a^1_{1})}^1\cdots  
  \oh{\psi}_{\sigma_{1}(a^{N_c}_{1})}^{N_c} 
  \right)\right\} \notag \\
  &\Bigg\{ \frac{1}{(Q!)^{N_c}}\sum_{\sigma^1 \cdots \sigma^{N_c}}
  \sgn(\sigma^1)\cdots \sgn(\sigma^{N_c}) \notag \\
  &\hspace{4cm}\left(\delta^{c^1_{\sigma^1(Q)}}_{Q}\cdots 
  \delta^{c^{N_c}_{\sigma^{N_c}(Q)}}_{Q} \right)\cdots 
  \left(\delta^{c^1_{\sigma^1(1)}}_{1}\cdots 
  \delta^{c^{N_c}_{\sigma^{N_c}(1)}}_{1} \right)
  \Bigg\} \notag \\
  &\left\{\int_{\text{U}(N_f)}d\mu(h_{-})
  \Gamma^{(a^1_{Q}\cdots a^{N_c}_{Q})\cdots(a^{1}_{1}\cdots
  a^{N_c}_{1})}_{(c^1_{Q}\cdots c^{N_c}_{Q})
  \cdots(c^{1}_{1} \cdots c^{N_c}_{1})}
  \left(\Gamma^{(b^1_{Q}\cdots b^{N_c}_{Q})\cdots(b^{1}_{1}
  \cdots b^{N_c}_{1})}_{(d^1_{Q} \cdots d^{N_c}_{Q})
  \cdots(d^{1}_{1}\cdots d^{N_c}_{1})}\right)^{*} 
  \right\} \notag \\
  &\Bigg\{ \frac{1}{(Q!)^{N_c}}\sum_{\rho^1 \cdots \rho^{N_c}}
  \sgn(\rho^1)\cdots \sgn(\rho^{N_c}) \notag \\
  &\hspace{4cm}\left(\delta^{d^1_{\rho^1(Q)}}_{Q}\cdots 
  \delta^{d^{N_c}_{\rho^{N_c}(Q)}}_{Q} \right)\cdots 
  \left(\delta^{d^1_{\rho^1(1)}}_{1}\cdots 
  \delta^{d^{N_c}_{\rho^{N_c}(1)}}_{1} \right)
  \Bigg\} \notag \\
  &\left\{\frac{1}{(N_c!)^Q}\sum_{\rho_{1}\cdots\rho_{Q}}
  \left(\wh{\psi}_{\rho_{1}(b^{N_c}_{1})}^{N_c}\cdots 
  \wh{\psi}_{\rho_{1}(b^1_{1})}^1 \right)
  \cdots\left(\wh{\psi}_{\rho_{Q}(b^{N_c}_{Q})}^{N_c}\cdots 
  \wh{\psi}_{\rho_{Q}(b^1_{Q})}^{1}
  \right)\right\}.
  \label{eq:long-chi-q}
\end{align}
Here, $\sigma$ and $\rho$ denote permutations, where the upper (lower)
indices correspond to permutations of the color (flavor) indices.
As the hypermatrix $\Gamma$ is simply the representation matrix
corresponding to the irreducible 
representation of U($N_f$) 
shown in Fig.~\ref{fig:u-n-young-graph},
we may use the group theoretic result that for irreducible
unitary representations $\Gamma^{k}$ and $\Gamma^{k'}$ of the compact
Lie group $G$,
\begin{equation}
\label{eq:ortho_rel}
  \int_{G}d\mu(g)\left(\Gamma^{k}(g)_{ij} \right)^{*}
  \Gamma^{k'}(g)_{i'j'}=\frac{1}{\lambda_k}\delta_{ii'}\delta_{jj'}\delta_{kk'}
 \int_{G}d\mu(g)\:.
\end{equation}
Here, $\lambda_k$ is the dimension of the representation $\Gamma^k$ of
$G$. We normalize the group volume of U($N_f$) to unity.

For convenience, we define $\lambda_n^N$ to be the dimension of a
representation of U($N$) which has a rectangular Young diagram with
$n$ rows and $N_c$ columns.  It is given by Weyl's formula,
\begin{equation}
\label{eq:dim-q}
  \lambda_n^N = \frac{\prod_{1\leq i<j\leq N}(\eta_i - \eta_j-i+j)}{
              (N-1)!(N-2)!\cdots 0!}\:,
\end{equation}
where $i$ and $j$ enumerate the rows of the Young diagram, and
$\eta_i$ is the length of the $i$-th row. Thus, $\eta_i = N_c$ if
$1\leq i\leq n$ and $\eta_i = 0$ otherwise.  For the present
application of Eq.~\eqref{eq:ortho_rel}, the $\lambda_k$ are simply
given by $\lambda_Q^{N_f}$.

It is now a matter of unraveling the expression for $\chi_Q$ in
Eq.~\eqref{eq:long-chi-q}. An important part of this expression is
\begin{align}
  \frac{1}{(N_c!)^{2Q}}&\sum_{\substack{\sigma_1\cdots\sigma_Q \\
                                             \rho_1\cdots\rho_Q}}
\left[(\oh{\psi}^1_{\sigma_Q(a_Q^1)}\cdots 
       \oh{\psi}^{N_c}_{\sigma_Q(a_Q^{N_c})}) \cdots
      (\oh{\psi}^1_{\sigma_1(a_1^1)}\cdots
       \oh{\psi}^{N_c}_{\sigma_1(a_1^{N_c})}) \right]
\left[(\delta^{a_Q^1}_{b_Q^1}\cdots 
       \delta^{a_Q^{N_c}}_{b_Q^{N_c}})\right. \notag \\
&\quad\cdots
\left.(\delta^{a_1^1}_{b_1^1}\cdots
       \delta^{a_1^{N_c}}_{b_1^{N_c}})\right]
\left[(\wh{\psi}^{N_c}_{\rho_1(b_1^{N_c})}\cdots
       \wh{\psi}^1_{\rho_1(b_1^1)})\cdots
      (\wh{\psi}^{N_c}_{\rho_Q(b_Q^{N_c})}\cdots
       \wh{\psi}^1_{\rho_Q(b_Q^{1})})\right] \notag \\
=\frac{1}{(N_c!)^{2Q}}&\sum_{\substack{\sigma_1\cdots\sigma_Q \\
                                             \rho_1\cdots\rho_Q}}
\sgn(\sigma_1)\cdots \sgn(\sigma_Q)\sgn(\rho_1)\cdots\sgn(\rho_Q)
 \left[(\oh{\psi}_{a_Q^1}^{\sigma_Q^{-1}(1)}\cdots
       \oh{\psi}_{a_Q^{N_c}}^{\sigma_Q^{-1}(N_c)})\right.\notag \\
\cdots
      &(\oh{\psi}_{a_1^1}^{\sigma_1^{-1}(1)}\cdots
 \left.\oh{\psi}_{a_1^{N_c}}^{\sigma_1^{-1}(N_c)})\right]
\left[(\wh{\psi}_{a_1^{N_c}}^{\rho_1^{-1}(N_c)}\cdots
       \wh{\psi}_{a_1^1}^{\rho_1^{-1}(1)})\cdots
      (\wh{\psi}_{a_Q^{N_c}}^{\rho_Q^{-1}(N_c)}\cdots
       \wh{\psi}_{a_Q^1}^{\rho_Q^{-1}(1)})\right] \notag \\
=\frac{1}{(N_c!)^{2Q}}&\left( \sum_{\sigma,\rho}\sgn(\sigma)\sgn(\rho)
\prod_{i=1}^{N_c}\oh{\psi}_a^{\sigma(i)}\wh{\psi}_a^{\rho(i)}\right)^Q.
\label{eq:unravel}
\end{align}
The term in parentheses in the last line of Eq.~\eqref{eq:unravel} is
the Laplace expansion for a determinant (more precisely, $N_c!$ copies
thereof). Thus, we obtain
\begin{equation}
\chi_Q = \frac{C_Q}{\lambda_Q^{N_f} C_0(Q!)^{N_c}}\left(
\frac{\det(\mathcal{M})}{N_c!}\right)^Q,
\end{equation}
where the matrix $\mathcal{M}$ has entries $\mathcal{M}^{ij}=\bar{\psi}_a^i
(1+ZZ^{\dag})_{ab}\psi_b^j$. 

For $Q<0$, the argument is analogous, and we find
\begin{equation}
  \chi_Q=\frac{C_{|Q|}}{\lambda_{|Q|}^{N_f}C_0(|Q|!)^{N_c}}
         \left(\frac{\det(\mathcal{N})}{N_c!}\right)^{|Q|}.
\end{equation}
The matrix $\mathcal{N}$ has entries 
$\mathcal{N}^{ij}=\bar{\varphi}_a^i (1+Z^{\dag}Z)_{ab}\varphi_b^j$.

Finally, let us find the constants $C_Q$. To this end,
we evaluate the matrix element
\begin{equation}
1 = \braopket{\psi_Q}{\1_Q}{\psi_Q} 
=C_Q \int_{\text{U}(2N_f)} d\mu(g) 
\braopket{\psi_Q}{T_g^{\phantom{1}}}{\psi_Q}
\braopket{\psi_Q}{T_g^{-1}}{\psi_Q}\:.
\end{equation}
Note that $\braopket{\psi_Q}{T_g}{\psi_Q}$ is simply a diagonal
element of the representation matrix $\Gamma^{Q}(g)$ of the
representation $Q$ of the group U($2N_f$) in
Fig.~\ref{fig:young-graph}(c). Similarly,
$\braopket{\psi_Q}{T_g^{-1}}{\psi_Q}$ is the complex conjugate of this
matrix element. Hence, we can once again use Eq.~\eqref{eq:ortho_rel}
together with Weyl's formula \eqref{eq:dim-q} for the dimension of
the representation $Q$ of U($2N_f$) shown in
Fig.~\ref{fig:young-graph}(c) to obtain
\begin{equation}
\label{eq:proj_const}
C_Q = \lambda_{N_f+Q}^{2N_f}\:.
\end{equation}
In particular, the overall normalization constant in
Eq.~\eqref{eq:col_flav} is
\begin{equation}
  \label{eq:norm2}
  C_0=\prod_{n=0}^{N_f-1}\frac{n!(N_c+N_f+n)!}{(N_c+n)!(N_f+n)!}\:.
\end{equation}
Note that $C_Q=C_{-Q}$.  Thus, the constants $\mathcal{C}_Q$ which
appear in Eq.~\eqref{eq:chi_q} are given by
\begin{align}
  \mathcal{C}_Q&=\frac1{(Q!)^{N_c}(N_c!)^Q}
  \frac{\lambda_{N_f+Q}^{2N_f}}{\lambda_Q^{N_f} \lambda_{N_f}^{2N_f}}\:. 
\end{align}
Using Eq.~\eqref{eq:dim-q}, a number of cancellations then leads to
the result quoted in Eq.~\eqref{eq:CQ}.  This completes the proof.

\section{Conclusions and outlook}
\label{sec:concl}

We have generalized Zirnbauer's color-flavor transformation in the
fermionic sector to the case of gauge group SU($N_c$).  This
transformation is expected to have a number of interesting
applications in lattice gauge theories with gauge group SU($N_c$).  In
fact, a first application has already appeared in
Ref.~\cite{Budczies:2000qs} where a low-energy effective theory, valid
in the strong-coupling limit, was derived in the vacuum ($Q=0$) and
one-baryon ($Q=1$) sectors.

There are a number of obvious open problems.  The most difficult one
appears to be the inclusion of the gauge action beyond the
strong-coupling limit.  Another issue is the problem of chiral symmetry
on the lattice which has recently been solved in a variety of ways
that can all be traced back to the Ginsparg-Wilson relation
\cite{Gins82}.  From the point of view of the color-flavor
transformation, is seems easiest to address this problem in the
domain-wall fermion approach \cite{DWF}.

Finally, it would be interesting to extend the color-flavor
transformation with gauge group SU($N_c$) to the bosonic sector and to
the supersymmetric case where the tensors $\psi$ and $\varphi$ contain
both bosonic and fermionic variables.  This results in some
complications since the sum over $Q$ in Eq.~\eqref{eq:sumproj} then
extends from $-\infty$ to $\infty$.  Work in this direction is in
progress.

\section*{Acknowledgements}
This work was supported in part by the U.S.\ Department of Energy
under contract no.\ DE-FG02-91ER40608.  We thank S. Shatashvili for
interesting conversations.

\begin{appendix}
\section{Appendix}
In this appendix we give some simple examples to illustrate 
the color-flavor transformation.
 
{\bf Example 1}\hspace{0.5cm} $N_c=1,\; N_f=1.$ \\
This is the simplest case, and it was already mentioned in
Ref.~\cite{Budczies:2000qs}. Integration over SU(1) simply amounts to
the evaluation of the integrand at unity, and so we have
\begin{equation}
\int_{\text{SU}(1)}dU \exp(\bar{\psi}U\psi+\bar{\varphi}U^{*}\varphi)
=\exp(\bar{\psi}\psi+\bar{\varphi}\varphi) 
=1+\bar{\psi}\psi+\bar{\varphi}\varphi+\bar{\psi}\psi
\bar{\varphi}\varphi\:.
\label{eq:left11}
\end{equation}
On the flavor side of the transformation, we have the integral over
the complex number $z=x+iy$,
\begin{equation}
\label{eq:right11}
\frac{2}{\pi}\int \frac{dx dy}{(1+zz^{*})^3}
\exp(\bar{\psi}z\varphi-\bar{\varphi}z^{*}\psi)
(\chi_0+\chi_1)\:.
\end{equation}
From Eq.~\eqref{eq:chi_q}, we find that $\chi_0 = 1$
and $\chi_1 = (1/2)(1+zz^{*})(\bar{\psi}\psi+\bar{\varphi}\varphi)$.
It is straightforward to expand the exponential and evaluate the 
integral in Eq.~\eqref{eq:right11} and to show that the resulting 
expression is equal to the right-hand side of Eq.~\eqref{eq:left11}.

{\bf Example 2}\hspace{0.5cm} $N_c=2,\; N_f=1.$\\
The integral over SU(2) color matrices can be evaluated by
parameterizing them as
\begin{equation}
U=\mat{e^{i\lambda}\cos\theta}{-e^{i\eta}\sin\theta}
      {e^{-i\eta}\sin\theta}{e^{-i\lambda}\cos\theta}\quad\text{with}
\quad 0\leq\theta\leq\frac{\pi}{2};\;
0\leq \lambda,\eta\leq 2\pi\:.
\end{equation}
The corresponding Haar measure is $dU=(1/2\pi^2)\sin\theta\cos\theta
d\theta d\lambda d\eta$.  We thus obtain
\begin{align}
\int_{\text{SU}(2)}dU \exp&\left(\bar{\psi}^i U^{ij}\psi^j
+\bar{\varphi}^i U^{\dagger ij}\varphi^j \right) 
=1+\bar{\psi}^1\psi^1\bar{\psi}^2\psi^2
+\bar{\varphi}^1\varphi^1\bar{\varphi}^2\varphi^2 \notag \\
&+\frac{1}{2}\left(\bar{\psi}^1\psi^1\bar{\varphi}^1\varphi^1
                 +\bar{\psi}^1\psi^2\bar{\varphi}^2\varphi^1
                 +\bar{\psi}^2\psi^1\bar{\varphi}^1\varphi^2
                 +\bar{\psi}^2\psi^2\bar{\varphi}^2\varphi^2\right).
\label{eq:left21}
\end{align}

For the flavor integral, we have (again with $z=x+iy$)
\begin{equation}
\frac{3}{\pi}\int \frac{dx dy}{(1+zz^{*})^4}
\exp\left(\bar{\psi}^i z\varphi^i -\bar{\varphi}^i z^{*}\psi^i \right)
(\chi_0+\chi_1)\:.
\end{equation}
Here, $\chi_0=1$ and $\chi_1 = (1/3)(1+zz^{*})^2 (\bar{\psi}^1 \psi^1
\bar{\psi}^2 \psi^2
+\bar{\varphi}^1\varphi^1\bar{\varphi}^2\varphi^2)$. Again, it is
straightforward to expand the exponential and evaluate the integral.
The result is equal to the right hand side of Eq.~\eqref{eq:left21}.

{\bf Example 3}\hspace{0.5cm} $N_c=1,\; N_f=2.$\\
In this case, as in Example 1, integration over SU(1) is trivial, and
we have
\begin{align}
\int_{\text{SU}(1)} dU& \exp(\bar{\psi}_a U \psi_a 
+\bar{\varphi}_a U^{*} \varphi_a)=
\exp(\bar{\psi}_a \psi_a +\bar{\varphi}_a \varphi_a) \notag\\
= 1&+\bar{\psi}_a\psi_a+\bar{\varphi}_a \varphi_a 
+\bar{\psi}_a\psi_a\bar{\varphi}_b \varphi_b
+\bar{\psi}_1\psi_1\bar{\psi}_2\psi_2 
+\bar{\varphi}_1\varphi_1\bar{\varphi}_2\varphi_2\notag \\
&+\bar{\psi}_a\psi_a\bar{\varphi}_1\varphi_1\bar{\varphi}_2\varphi_2
+\bar{\psi}_1\psi_1\bar{\psi}_2\psi_2\bar{\varphi}_a\varphi_a
+\bar{\psi}_1\psi_1\bar{\psi}_2\psi_2\bar{\varphi}_1\varphi_1
\bar{\varphi}_2\varphi_2\:.
\label{eq:left12}
\end{align}
Note that repeated indices are summed over. 
The corresponding flavor integral is
\begin{equation}
\frac{72}{\pi^4}\int\frac{dZdZ^{\dag}}{\det(1+ZZ^{\dag})^5}
\exp\left(\bar{\psi}_a Z_{ab}\varphi_b-\bar{\varphi}_a
Z^{\dag}_{ab}\psi_b \right)(\chi_0+\chi_1+\chi_2)\:,
\label{eq:right12}
\end{equation}
where $\chi_0=1$, $\chi_1=(1/3)(\bar{\psi}_a(1+ZZ^{\dag})_{ab}\psi_b
+\bar{\varphi}_a(1+Z^{\dag}Z)_{ab}\varphi_b)$, and $\chi_2 =
(1/6)\det(1+ZZ^{\dag})(\bar{\psi}_1\psi_1\bar{\psi}_2\psi_2
+\bar{\varphi}_1\varphi_1\bar{\varphi}_2\varphi_2)$.  This integral
can be evaluated using the singular value decomposition of $Z$ given
by $Z=U^{\dag}\Lambda V$. Here, $U\in U(2)$, $V\in U(2)/[U(1)\times
U(1)]$, and $\Lambda = \diag(\lambda_1, \lambda_2)$, where
$\lambda_1,\lambda_2\geq 0$. In these coordinates,
$\det(1+ZZ^\dagger)=\det(1+\Lambda^2)=(1+\lambda_1^2)(1+\lambda_2^2)$.
The Jacobian of the coordinate transformation is
$J=\lambda_1\lambda_2(\lambda_1^2-\lambda_2^2)^2$.  The measure of
integration is thus obtained using
\begin{equation}
dZ dZ^{\dag} = dU \: dV \: d\lambda_1\:
d\lambda_2\: \lambda_1\lambda_2(\lambda_1^2-\lambda_2^2)^2\:.
\end{equation}
From our normalization of the measure it follows that $\int
dU\:dV=2\pi^4$.  

We now expand the exponential in Eq.~\eqref{eq:right12} and drop all
terms for which the number of elements of $U$ is not equal to those of
$U^\dagger$, since such terms vanish when integrated over $U$.  We
also make repeated use of the identity $\int dU
U_{ij}^{\dag}U_{kl}=(1/2)\delta_{il}\delta_{jk}\int dU$.  Next, we
perform the integrals over the $\lambda_a$ and use the unitarity of
$U$ and $V$.  This eliminates all dependence on $U$ and $V$ from the
integrand.  After performing the trivial integration over $U$ and $V$,
one obtains the same result as in Eq.~\eqref{eq:left12}.
\end{appendix}


\end{document}